\documentclass{PoS}

\title{Prospects On Testing Lorentz Invariance Violation With The 
Cherenkov Telescope Array}

\ShortTitle{Testing LIV with CTA}

\author{\speaker{M.~K.~Daniel}\\
        Dept. of Physics, University of Liverpool, Liverpool, L69 7ZE. U.K.\\
        E-mail: \email{michael.daniel@liverpool.ac.uk}}

\author{D.~Emmanoulopoulos\\
        Dept. of Physics \& Astronomy, University of Southampton, Southampton, S017 1BJ. U.K.\\
        E-mail: \email{d.emmanoulopoulos@soton.ac.uk}}

\author{M.~Fairbairn\\
        Kings College London, London, WC2R 2LS. U.K.\\
        E-mail: \email{malcolm.fairbairn@kcl.ac.uk}}

\author{N.~Otte\\
        School of Physics \& Center for Relativistic Astrophysics, Georgia Institute of Technology, Atlanta, Georgia. USA\\
        E-mail: \email{otte@gatech.edu}}

\author{for the CTA Consortium\\
        https://www.cta-observatory.org/
}

\abstract{The assumption of Lorentz invariance is one of the founding 
principles of modern physics and violation of that would have 
deep consequences to our understanding of the universe. Potential 
signatures of such a violation could range from energy dependent 
dispersion introduced into a light curve to a change in the 
photon-photon pair production threshold that changes the expected 
opacity of the universe. Astronomical sources of Very High Energy 
(VHE) photons can be used as test beams to probe fundamental physics 
phenomena, however, such effects would likely be small and need to 
be disentangled from intrinsic source physics processes. The 
Cherenkov Telescope Array (CTA) will be the next generation ground 
based observatory of VHE photons. It will have improved flux 
sensitivity, a lower energy threshold (tens of GeV), broader energy 
coverage (nearly 5 decades) and improved energy resolution (better 
than 10\% over much of the energy range) compared to current 
facilities in addition to excellent time resolution for short 
timescale and rapidly varying phenomena. The expected sensitivity of 
this facility leads to us to examine in this contribution the kinds 
of limits to Lorentz Invariance Violation (LIV) that we could expect 
to obtain on VHE observations of Active Galactic Nuclei (AGN), Gamma 
Ray Bursts (GRBs) and pulsars with CTA. With a statistical sample 
and wide variety of sources CTA has the potential to set model 
independent limits.
}

\FullConference{The 34th International Cosmic Ray Conference,\\
		30 July- 6 August, 2015\\
		The Hague, The Netherlands}

\begin{document}

\section{Introduction}
\label{sec:liv-intro}
Lorentz invariance (LI) is one of the founding principles of the Special Relativity theory of Modern Physics. However it has long been understood that attempting to unify General Relativity (GR) with that other success of Modern Physics, Quantum Mechanics (QM), can in turn lead to deviations from Lorentz symmetry when describing spacetime structure in terms of finite quanta rather than as a continuous lightcone in Minkowski spacetime (\cite{amc2013, ell2013} and references therein). If LI is only an approximate symmetry of local spacetime then it is likely modified at some scale outside our realm of experience, the Planck scale ($E_\mathrm{QG} \approx E_\mathrm{Pl} \simeq 10^{19}$\,GeV) being a natural one to hypothesise. 

Whilst many Quantum Gravity (QG) models have been posited, because the scale of Lorentz invariance violation (LIV) is likely to be so far beyond anything that is feasibly accessible in the lab. any consequent effect on the observable world would be so correspondingly small that it can be treated perturbatively and approximated by a dispersion measure that is a simple Taylor expansion 
\begin{equation}
\label{eq:perturbation}
c^2 p^2 = E_{\gamma}^2\sum_{\alpha}\pm\xi_{\alpha}(E_{\gamma}^{\alpha}/E_{\rm{QG}}^{\alpha})
\end{equation}
where $c$ is the speed of light, $p$ the momentum, $E_{\gamma}$ the energy and $\xi_{\alpha}$ is the correction factor, with the leading linear ($\alpha=1$) and quadratic ($\alpha=2$) terms being those of the most interest. In the linear case, it has been shown that CPT can be violated in effective field theory \cite{ott2013}; however, if CPT is preserved and LI violated it is the quadratic term that would dominate. A positive correction term represents a subluminal change and a negative a superluminal one. This results in there being many QG models that lead to a vacuum velocity of light that is energy dependent. 

The most energetic photons recorded are from astrophysical sources and have energies of $\sim$ tens of TeV; for $E_{\gamma} \sim 1$\,TeV the correction to the speed of light due to Planck scale linear quantum gravity would be of order $10^{-15}$c. The infinitesimal magnitude of the signature at accessible energy ranges means that these searches require extremely sensitive measurements. Usefully, the minuscule corrections are cumulative and so can become a measurable dispersion when photons travel astronomical distances; although the magnitude of the time delays expected are still only $\delta t \leq 10$\,s/TeV/Gpc for a linear term Planck scale QG.

For measuring dispersion due to LIV there are three criteria that an ideal probe should meet:
\begin{itemize}
 \item emit very high energy photons, 
 \item be very distant, 
 \item exhibit variability with good statistics
\end{itemize}
Unfortunately some of these are mutually exclusive, for example very high energy photons will be attenuated by $\gamma + \gamma \rightarrow \mathrm{e}^{+} + \mathrm{e}^{-}$ pair production on the diffuse extragalactic background light, thereby limiting the distance to which these sources will have a detectable, time resolved signal.

\section{Cherenkov Telescope Array}
\label{sec:liv-cta}
\begin{figure}[htbp]
\begin{center}
\resizebox{0.5\textwidth}{!}{\includegraphics{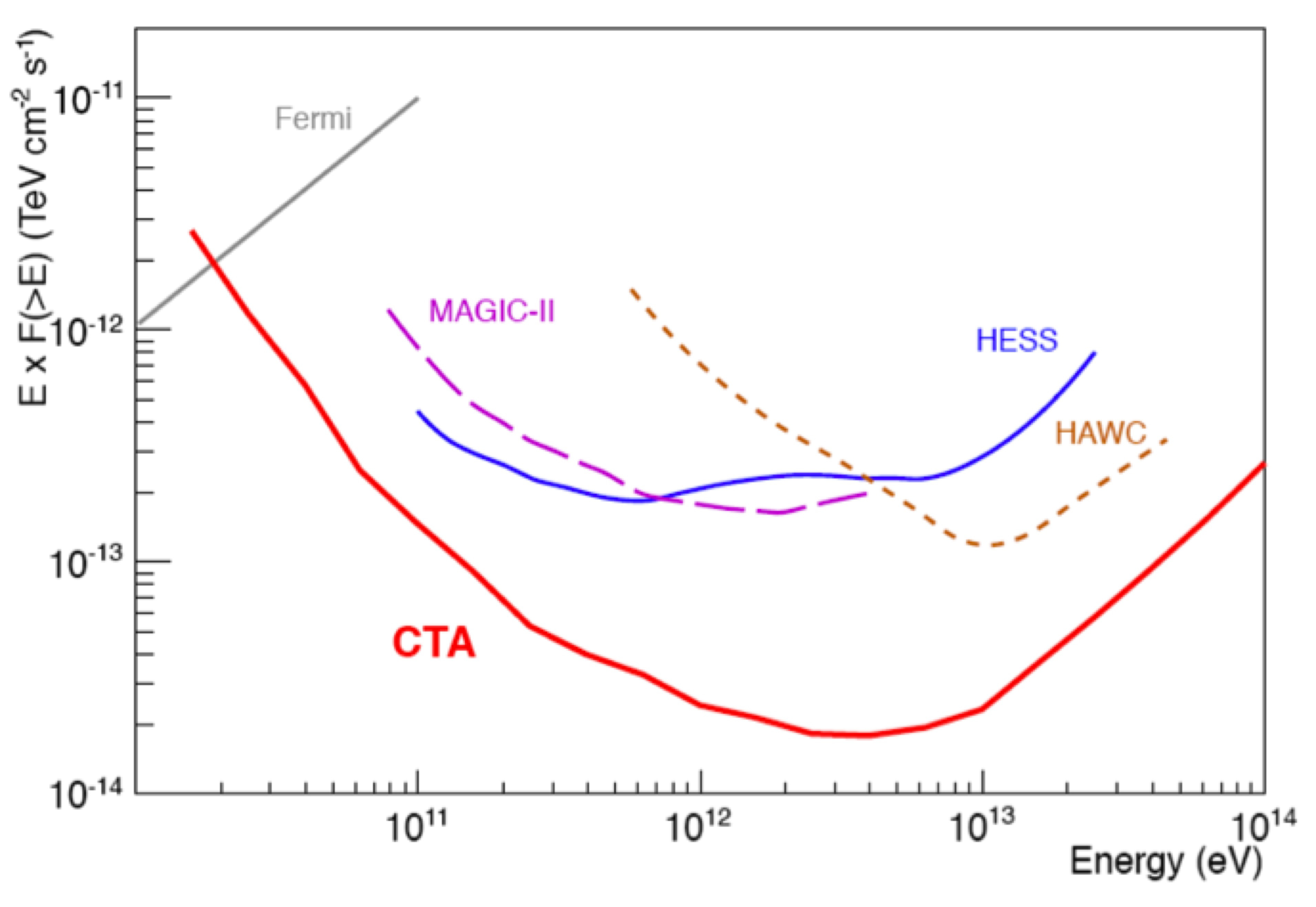}}
\caption[]
        {Integral Sensitivity of CTA in comparison to currently operating facilities.}
\label{fig:cta-sens}
\end{center} 
\end{figure}

The Cherenkov Telescope Array (CTA)\footnote{www.cta-observatory.org} \cite{CTA} will be the next generation imaging atmospheric Cherenkov telescope (IACT) facility, providing the community with an open-access observatory for the observation of gamma rays with energies from a few tens of GeV to hundreds of TeV with unprecedented sensitivity and angular and energy resolution. To achieve the large dynamic range in energy coverage the array will be comprised of multiple telescope sizes: 23\,m diameter large size telescopes provide a low energy threshold and fast slewing for objects like gamma ray bursts (GRBs); 12\,m diameter medium size telescopes provide an order of magnitude improvement in flux sensitivity ideal for surveys in the 0.1 to 1\,TeV region; and 4\,m diameter small sized telescopes (SSTs) will examine the highest energy region ($E \geq 10$\,TeV) ideal for examining cut-offs to the spectrum. Whole sky coverage will be achieved by operating at sites in both hemispheres. Figure~\ref{fig:cta-sens} shows the anticipated performance of CTA in comparison to current facilities.

\section{Time Of Flight Measurements}
\label{sec:liv-tof}

To search for a difference in the arrival times of high energy photons as a signature of quantum gravity effects was first suggested for gamma-rays bursts \cite{amc1998}, but has also enjoyed varying levels of success in the light curves of active galactic nuclei and pulsars, too. In order to disentangle intrinsic source effects, e.g. due to particle acceleration, from an externally induced dispersion will require measurements from a variety of sources and source types.

\subsection{Gamma ray bursts}
Gamma ray bursts (GRBs) show the fastest variability, where short duration GRBs have a duration of $<2$\,s, and have the furthest distance of known $\gamma$-ray sources (out to at least $z \sim 1$), but not the highest energy photons from known astrophysical sources. Observed by the Fermi-LAT satellite~\cite{FermiSpecs}, GRB\,090510 has already provided some of the most constraining limits yet, with a limit above the Planck scale for a linear term LIV induced dispersion \cite{FermiLimits, vas2013} using photons above 10\,GeV. 
CTA will have an effective collection area that is $\geq 10^{4}$ that of Fermi, so if a similar GRB is detected the photon statistics at the highest energies would increase significantly making for even more robust and constraining limits. The relatively small field of view of an IACT make it challenging to catch GRBs serendipitously at the onset phase and so a divergent pointing mode in survey observations would be the most likely observing strategy to successfully catch a GRB at its most relevant light curve stage \cite{ctaGRB}. GRBs also suffer in that they are unpredictable in location and distance, so it is challenging to build up statistics with them as individual sources with a pointed instrument.

\subsection{Active Galactic Nuclei}
\begin{figure}[htbp]
\begin{center}
\resizebox{0.5\textwidth}{!}{\includegraphics{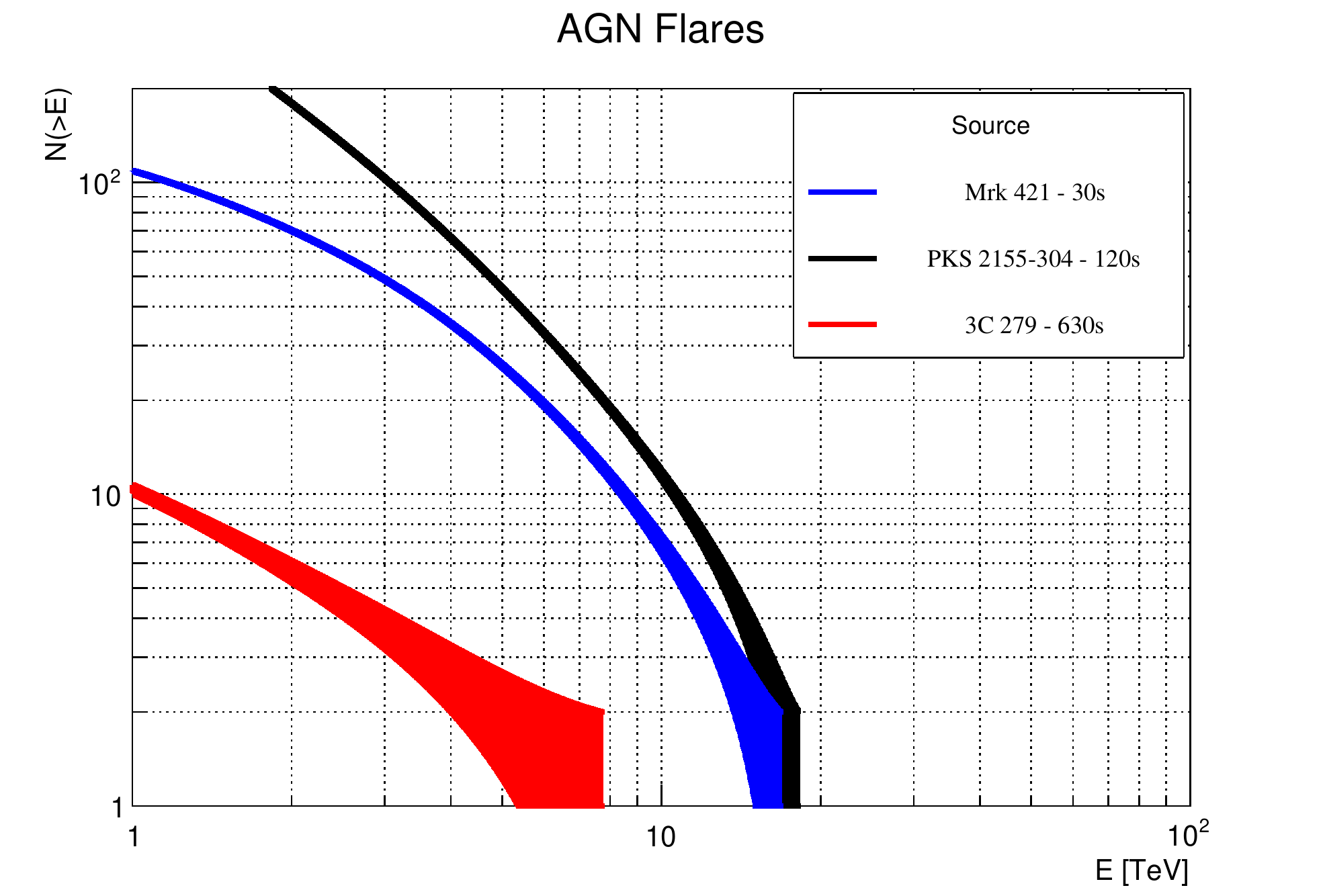}}
\caption[]
        {The number of photons CTA expects to see above a given energy in a given time extrapolating for the highest flux states for the AGN Mrk\,421 (blue line) in 30\,s; PKS\,2155-304 (black line) in 120\,s; and 3C\,279 (red line) in 630\,s.}
\label{fig:AGN}
\end{center} 
\end{figure}

The jets from active galactic nuclei (AGN) make for even more energetic sources of photons than GRBs, with the advantage that they also have known positions that enable them to be monitored for activity, but at the expense of longer ($\sim$ minute) scale variability features in the light curve. Current LIV limits are just below the Planck scale on the linear term with current generation instruments observations of AGN flares \cite{xcol2155, likelihood2155, magicqg}. To further constrain the limits will require the CTA to detect significant numbers of higher energy photons and/or faster time variability from AGN flare observations. Following the method of \cite{bar2012} shows that a minimum of 10 photons are required in the high energy light curve component of a flare that is no longer in duration than 3 times the expected level of dispersion to perform a test; and to improve on current limits would require a ten-fold increase in the energy of photons tested, i.e. an enhanced collection area for photons above 10\,TeV. Taking the highly active flaring states and extrapolating the spectra for the AGN Mrk\,421~\cite{Mrk421a}, PKS\,2155-304~\cite{PKS2155} and 3C\,279~\cite{3C279} we plot in figure~\ref{fig:AGN} the number of expected photons above a given energy based on the CTA expected performance. For PKS\,2155-304 it appears very achievable to obtain at least 10 photons above 10\,TeV in a 120\,s flare, which is within a factor 2 of the currently measured most extreme flaring activity and also close within reach of Mrk\,421 in 30\,s which is regularly seen in a high flaring state (having exceeded 10 Crab on at least 3 occassions in 14 years of monitoring~\cite{Mrk421b}). So if these extreme conditions were repeated, or exceeded, for CTA observations the prospects for improving the current LIV limits seem to be good to match GRB limits on the linear term and likely exceed them on the quadratic term.

Even if the AGN light curve does not have sufficiently rapid features to determine dispersion for any single flaring episode, a LIV induced dispersion will mean that higher energy photons will always arrive shifted with respect to lower energy ones in the light curve. The accumulation of long term monitoring data means that we can still potentially determine time delays at high confidence, e.g. through the use of cross-power spectral analysis methods \cite{dor2013}. The underlying rapid varying features monitored over long periods serving to further increase the chance of detection. This will be the first time that routine AGN observations, i.e. not on exceptional flux levels, will provide us with such LIV constraints.

\subsection{Pulsars}
When it comes to testing the quadratic term for LIV, having a very high energy component compensates for a lack of distance. Pulsars represent a very fast varying, relatively well understood source population with very different intrinsic source physics processes to GRBs and AGN. The Crab pulsar has a pulsed VHE component to its spectrum up to hundreds of GeV and little evidence of a cut-off (within event statistics) \cite{MAGICCrabPulsar}. Whilst the LIV limits from current generation instruments are presently inferior to those from AGN and GRBs \cite{zit2013}, CTA observations for 50 hours with a cut-off to the spectrum that is $>1$\,TeV would result in limits that are competitive with the current generation limits from the AGN PKS\,2155-304. An inverse Compton component to the pulsar spectrum could be present that would generate VHE emission \cite{Takata2006, Hirotani2007}. Assuming a millisecond pulsar observed above 1\,TeV with CTA if the pulsed flux is 10\% the pulsar flux at 100\,GeV then in 100 hours of observations limits at the Planck energy scale in the linear term and $\simeq 10^{11}$\,GeV level in the quadratic term are potentially achievable. This opens up the possibility of pulsars providing comparable limits to AGN/GRBs. As a pulsar has a very well measured light curve profile it also makes for an interesting source to test for any light curve broadening that might occur from a polarisation dependent superluminal correction (see e.g.~\cite{superluminal, IntegralLimits}), which would give this source class a distinct advantage over AGN/GRBs.


\section{Gamma Ray Horizon Measurements}
\label{sec:liv-horizon}
If LIV modifies the dispersion relation for $\gamma$ rays it could also affect the kinematics in the pair production process through the addition of an extra correction 
\begin{displaymath}
E^{2} = p^{2}c^{2} + m^{2}c^{4} \pm E^{2} \left ( \frac{E}{E_{QG}} \right )^{\alpha}
\end{displaymath}
changing the cross-section for interaction that attenuates the VHE signal as it travels through the diffuse extragalactic background radiation filling the universe. This would then allow VHE photons, potentially up to hundreds of TeV, to be detected that would not normally be expected in deep observations of suitably hard spectrum distant AGN/GRBs \cite{kif1999, fai2014, Biteau2015}  that are used to constrain the level of extragalactic background light (EBL) in the hard to directly measure ultraviolet-infrared region or on the cosmic microwave backround (CMB). Any photons detected at or above the energies where attenuation is expected on the CMB would be an unambiguous signal for new physics since the diffuse emission spectrum is extremely well known there and the horizon extremely small, due to the CMB being 4 orders of magnitude larger than the EBL. As the energy region to target is for photons above 10\,TeV, it is important that the SST component of the array be included in deep observations of AGN. Appropriate source selection is paramount, ideally an object with any cut-off to the intrinsic source spectrum to be higher than 10\,TeV, meaning that sources with a harder spectrum, like AP\,Lib, that may be relatively faint below 1\,TeV can be more suitable than bright ones, such as PKS\,2155-304.  Table~\ref{tab:horizon} shows if we take a generic source at the $\simeq 2\%$ Crab level of emission with a spectrum of $1.12 \times 10^{-12} E^{-2.72} \exp(-E/10\mathrm{TeV})$\,photons\,cm$^{-2}$\,s$^{-1}$ that very competitive limits can be made, particularly on the quadratic term. Interestingly, the region in redshift where the best limits can be found is slightly further away for the quadratic term than the linear one, which at first seems slightly counter-intuitive given that in the time-of-flight measurements the distance requirement is lessened greatly by the square in the energy term.

\begin{table}[htbp]
\begin{center}
\begin{tabular}{|c|c|c|}   
\hline
redshift & linear & quadratic \\
z & $[$GeV$]$ & $[$GeV$]$ \\
\hline
0.05 & $ 2.03 \times 10^{18} $ & $ 1.61 \times 10^{11} $ \\
0.10 & $ 1.43 \times 10^{18} $ & $ 2.21 \times 10^{11} $ \\
0.15 & $ 1.24 \times 10^{18} $ & $ 1.23 \times 10^{11} $ \\
\hline
\end{tabular}
\end{center}
\caption[]
        {$3\sigma$ limits on the QG energy scale for a source at different redshifts and intrinsic spectrum of 2\% Crab flux spectral index of -2.7 and an exponential cut-off at 10\,TeV.}
\label{tab:horizon}        
\end{table}

\section{Summary}

CTA will be able to perform tests for LIV both in terms of time-of-flight tests that look for an energy dependent dispersion and for changes to the pair-production interaction threshold that would shift the gamma-ray horizon. With an appropriate observing strategy and source selection, the potential is there to place limits on a linear correction term at the Planck scale and $>10^{11}$\,GeV on a quadratic term. 
By having sensitivity to test on a wide variety of sources as a function of redshift, on different source categories and different physical processes enables CTA to perform these tests in a model independent fashion for the first time in VHE observations.

We gratefully acknowledge support from the agencies and organizations listed under Funding Agencies at this website: http://www.cta-observatory.org/.

\bibliographystyle{JHEP}
\bibliography{mkd-liv}



\end{document}